\documentclass[aip,jcp,reprint]{revtex4-1}
\usepackage{epsfig}
\usepackage{color}
\usepackage{amsmath}
\usepackage{amsfonts}
\usepackage{natbib}
\usepackage{notes2bib}
\usepackage{physics}

\newcommand{\rb}{\mathbf{r}}
\renewcommand{\pb}{\mathbf{p}}
\newcommand{\mb}{\mathbf{m}}

\newcommand{\avg}[1]{\left<#1\right>}
\newcommand{\len}[1]{\left|#1\right|}
\newcommand{\brac}[1]{\left[#1\right]}
\newcommand{\curly}[1]{\left\{#1\right\}}
\newcommand{\para}[1]{\left(#1\right)}

\newcommand{\eg}{\emph{e.g.}}

\newcommand{\etal}{\emph{et al.}}

\newcommand{\kT}{\ensuremath{k_{\rm B}T}}

\newcommand{\Hb}{\ensuremath{\mathcal{H}}}
\newcommand{\Vb}{\ensuremath{\mathcal{V}}}
\newcommand{\Tb}{\ensuremath{\mathcal{T}}}
\newcommand{\Zb}{\ensuremath{\mathcal{Z}}}

\newcommand{\mxx}{\ensuremath{m_{xx}}}
\newcommand{\mxy}{\ensuremath{m_{xy}}}
\newcommand{\mxz}{\ensuremath{m_{xz}}}
\newcommand{\myy}{\ensuremath{m_{yy}}}
\newcommand{\myz}{\ensuremath{m_{yz}}}

\begin{document}



\title{Effective Mass Path Integral Simulations
of Quasiparticles in Condensed Phases}


\author{Richard C. Remsing}
\email[]{rick.remsing@rutgers.edu}
\affiliation{Department of Chemistry and Chemical Biology, Rutgers University, Piscataway, NJ 08854}

\author{Jefferson E. Bates}
\email[]{jeb@appstate.edu}
\affiliation{A. R. Smith Department of Chemistry and Fermentation Sciences, Appalachian State University, Boone, NC 28608}




\begin{abstract}
The quantum many-body problem in condensed phases is often 
simplified using a quasiparticle description,
such as effective mass theory for electron motion in a periodic solid. 
These approaches are often the basis 
for understanding many fundamental condensed phase processes,
including the molecular mechanisms underlying solar energy harvesting and photocatalysis.
Despite the importance of these effective particles,
there is still a need for computational
methods that can explore their behavior on chemically relevant length and
time scales.
This is especially true when the interactions
between the particles and their environment are important.
We introduce an approach for studying quasiparticles in condensed phases
by combining effective mass theory with the path integral treatment
of quantum particles. 
This framework incorporates the generally anisotropic electronic band structure
of materials into path integral simulation schemes to enable modeling
of quasiparticles in quantum confinement, for example. 
We demonstrate the utility of effective mass path integral simulations
by modeling an exciton in solid potassium chloride
and electron trapping by a sulfur vacancy
in monolayer molybdenum disulfide.
\end{abstract}


\maketitle

\raggedbottom


%
Electronic excitations in semiconducting materials form the foundation of many
areas of materials and energy sciences, including solar energy harvesting and conversion and nanoelectronics.
It is often advantageous to describe processes involving such excitations
within the language of quasiparticles, e.g. electrons and holes with effective masses or excitons~\cite{kittel1976introduction,anderson1997concepts,mahan2013many}. 
Due to the complexity of these descriptions, the theory and simulation of quasiparticles
is often limited to coarse-grained and continuum approaches or modeling small,
highly symmetric systems in quantum mechanical detail~\cite{Cho_2018,Berkelbach_Review,Berkelbach_2013,Hess_2005,WillardPerspective,Goodman:2020aa,Troisi_2011,Schleife_2016}. 
While these approaches are responsible for important advances in our understanding of exciton
physics and nanotechnology,
it is difficult for existing methodologies to describe sources of disorder (defects)
and large system sizes are often needed to properly model the effects of charge carriers on their surroundings
and vice versa. 
One promising approach for simulating quantum particles in complex environments
uses the path integral (PI) representation of quantum mechanics, in which a quantum
particle, such as an electron, can be represented as a classical ring polymer.
Given pseudopotentials to describe the interactions between a quantum particle and its (often classical)
environment, one can perform molecular dynamics (MD) simulations
of large systems for long times. 
However, it is difficult for straightforward PIMD simulations to describe phenomena such as quantum confinement
that manifest as a result of anisotropic electronic band structures, 
in addition to quasiparticles such as holes. 
In this work, we describe an approach that uses effective mass theory~\cite{Slater:PhysRev:1949,Adams:PhysRev:1952,Luttinger:PhysRev:1955,Wannier:PhysRev:1937,Peierls:ZFP:1933,Dresselhaus:1956}
to incorporate these aspects of anisotropic electronic band structures of materials
into the path integral representation of quantum mechanics~\cite{FeynmanPathIntegrals,Chandler:JCP:1981,Berne:AnnRevPhysChem:1986,Sprik:CompPhysRep:1988,ceperley1995path}
in order to model quantum charge carriers and their excited states (e.g. excitons)
in complex, atomistic environments. 
We demonstrate the utility of this effective mass path integral (EMPI) approach by modeling
excitons in crystalline potassium chloride and electron trapping
in a defective monolayer of molybdenum disulfide (MoS$_2$).
%


%
We consider a quantum particle described by a Hamiltonian $\Hb=\Tb+\Vb$, which consists of a kinetic term, $\Tb$,
and a potential term describing its interactions with the environment, $\Vb$.
The partition function for this particle can be written as
\begin{equation}
\Zb = \int d\rb_1 \mel{\rb_1}{e^{-\beta\Hb}}{\rb_1}.
\end{equation}
By applying the (symmetric) Trotter factorization~\cite{FeynmanPathIntegrals,Chandler:JCP:1981,Berne:AnnRevPhysChem:1986,Habershon:2013aa},
we can write $\Zb$ in a form that physically corresponds to a discretization of the (cyclic) quantum path
of the particle in imaginary (or Euclidean) time,
\begin{align}
\Zb &= \lim_{P\rightarrow\infty} \int d\rb_1 \int d\rb_2 \cdots \int d\rb_P \nonumber \\
&\bigg[ \mel{\rb_1}{e^{-\beta \Vb(\rb_1)/2P}e^{-\beta\Tb/P}e^{-\beta\Vb(\rb_2)/2P}}{\rb_2} \nonumber \\
&\cdots  \mel{\rb_P}{e^{-\beta \Vb(\rb_P)/2P}e^{-\beta\Tb/P}e^{-\beta\Vb(\rb_1)/2P}}{\rb_1} \bigg].
\end{align}
In practice, we use a finite number of discretizations, $P$, and the partition function and
equilibrium ensemble averages are exact in the limit $P\rightarrow\infty$, corresponding to a continuous path.
The potential term, $\Vb(\rb)$, can be readily evaluated, and
so we focus on rewriting the kinetic part of the partition function.
This arises from matrix elements of the form
\begin{equation}
\mel{\rb_i}{e^{-\beta\Tb/P}}{\rb_j}=\int d\pb \mel{\rb_i}{e^{-\beta\Tb/P}}{\pb} \braket{\pb}{\rb_j},
\end{equation}
which connect each discrete step in the imaginary time path of the particle; \eg \ step $i$ to $j$.

We now work within effective mass theory (EMT)
to include some aspects of electronic structure in our model
through the above matrix elements.
EMT prescribes an effective mass $m^*$ to charge carriers
which reflects the modification of their masses, from that of a free electron,
due to the interactions between the charge carriers 
and the static nuclei~\cite{Slater:PhysRev:1949,Adams:PhysRev:1952,Luttinger:PhysRev:1955,Wannier:PhysRev:1937,Peierls:ZFP:1933}. 
In order to model quantum particles in a classical bath using the path integral isomorphism,
we substitute the free masses of the quantum particles with those determined from EMT
to include the influence of the electronic response to the environment in effective classical models.
For highly symmetric materials, isotropic parabolic fits of the band structure may be sufficient,
such that a scalar effective mass $m^*$ can be assigned to the particle~\cite{shumway2004quantum,Bischak:2018aa}.
However, lower symmetry materials require that the $3\times3$ mass-tensor
\begin{align}
\left(\frac{1}{m}\right)_{\alpha\gamma} = \frac{1}{\hbar^2} \frac{\partial^2 E(\mathbf{k})}{\partial k_\alpha \partial k_\gamma} \, ;\, \alpha,\gamma \in \curly{x,y,z}
\label{eq:emc}
\end{align}
be computed,
leading to an inverse effective mass \emph{tensor} $\mb^{-1}$
for each quantum particle that can in principle be anisotropic.
Note that the effective mass tensor is symmetric; $m^{-1}_{\alpha\gamma}=m^{-1}_{\gamma\alpha}$.
We also drop the star for notational clarity.
In the context of EMT, the kinetic energy of the quantum particle is now
\begin{equation}
\Tb = \frac{1}{2} \pb^{\rm T} \cdot \mb^{-1} \cdot \pb,
\label{eq:qke}
\end{equation}
which reduces to $\Tb=p^2/2m$ in the limit of an isotropic, diagonal mass matrix.
We can then follow the typical evaluation of the partition function in the path integral isomorphism, but now with
Eq.~\ref{eq:qke} for the quantum kinetic energy.
The desired matrix element can be readily evaluated through Gaussian integration~\cite{chaikin2000principles} to yield
\begin{align}
\nonumber
\mel{\rb_i}{e^{-\beta\Tb/P}}{\rb_j} & =\para{\frac{P}{2 \pi \beta  \hbar^2}}^{3/2} \det\brac{\mathbf{m}^{-1}}^{-1/2} \\
&\quad \times \exp\curly{-\sum_{\alpha,\gamma} \frac{1}{2}\frac{P}{\hbar^2\beta} m_{\alpha\gamma} \alpha_{ij} \gamma_{ij}},
\end{align}
where $m_{\alpha\gamma}$ is the $\alpha,\gamma$ element of the inverse of the matrix $\mathbf{m}^{-1}$.
With this expression for the matrix elements,
the partition function is given by
\begin{align}
\nonumber
\Zb &= \lim_{P\rightarrow \infty}\para{\frac{P}{2\pi\beta \hbar^2}}^{3P/2} \det\brac{\mathbf{m}^{-1}}^{-P/2} \\
&\quad \times \int d\rb_1 \cdots \int d\rb_P e^{-\beta \Hb_P (\rb_1,\cdots,\rb_P)},
\end{align}
where $\Hb_P$ is the isomorphic Hamiltonian of a classical ring polymer with harmonic bonds between neighboring
beads of the polymer.
This isomorphic Hamiltonian is
\begin{align}
\Hb_P(\rb_1,\cdots,\rb_P)&= \frac{1}{P}\sum_{i=1}^P \bigg[ \Vb(\rb_i) \\
&+ \sum_{\alpha,\gamma} \frac{\kappa_{\alpha\gamma}}{2} (\alpha_i-\alpha_j)(\gamma_i-\gamma_j) \bigg],
\end{align}
where
\begin{equation}
\kappa_{\alpha\gamma} = \frac{P^2 m_{\alpha\gamma}}{\beta^2\hbar^2},
\end{equation}
and the matrix elements $m_{\alpha\gamma}$ can be determined by inverting $\mathbf{m}^{-1}$.
The harmonic bonds between neighboring beads of the ring polymer generally are not spherically symmetric,
but involve different spring constants, $\kappa_{\alpha\gamma}$, along each direction, as well as coupling between
the displacements in the Cartesian components, as demonstrated below.
In the limit of a diagonal $\mb^{-1}$, the ring polymer Hamiltonian becomes
\begin{align}
\Hb_P(\rb_1,\cdots,\rb_P)&=\frac{1}{P} \sum_{i=1}^P \bigg[\Vb(\rb_i) + \frac{\kappa_{xx}}{2}(x_i-x_j)^2 \\
&+\frac{\kappa_{yy}}{2}(y_i-y_j)^2+\frac{\kappa_{zz}}{2}(z_i-z_j)^2 \bigg].
\end{align}
In this formulation, the harmonic springs between neighboring beads of the ring polymer do not have spatially isotropic spring constants,
but are instead given by $\kappa_{\alpha\alpha} = P^2 m_{\alpha \alpha}/\beta^2\hbar^2$,
where $\alpha$ refers to a spatial coordinate.
This is particularly important in systems with reduced dimensionality.
For example, $m_{zz}>>m_{xx}\approx m_{yy}$ in monolayer MoS$_2$, as discussed below,
resulting in ring polymers that are confined essentially to two dimensions.
To illustrate a specific case where off-diagonal coupling is present,
we explicitly consider an effective mass tensor with $m_{xz}^{-1}=m_{zx}^{-1}=m_{yz}^{-1}=m_{zy}^{-1}=0$;
all other elements are non-zero.
In this case, 
\begin{align}
\Hb_P(\rb_1,\cdots,\rb_P)&=\frac{1}{P}\sum_{i=1}^P \bigg[ \Vb(\rb_i) + \frac{\kappa_{xx}}{2}(x_i-x_j)^2 \nonumber \\
&+\frac{\kappa_{yy}}{2}(y_i-y_j)^2+\frac{\kappa_{zz}}{2}(z_i-z_j)^2 \nonumber \\
&- \kappa_{xy} (x_i-x_j)(y_i-y_j)  \bigg].
\label{eq:offdiag}
\end{align}
with the spring constants
\begin{equation}
\kappa_{\alpha \alpha} = \frac{P^2 m_{\alpha \alpha}}{\beta^2\hbar^2 \mu_{xy}}
\end{equation}
and
\begin{equation}
\kappa_{xy} = \frac{P^2 \mxx\myy}{\beta^2\hbar^2 \mxy \mu_{\rm xy}},
\end{equation}
where $\mu_{xy} = \para{1-\mxx \myy/\mxy^2}$.
The analogous Hamiltonian when $\mxz^{-1}$ or $\myz^{-1}$ is the only non-zero off-diagonal
element can be readily obtained by permuting the relevant indices in Eq.~\ref{eq:offdiag}.
Inclusion of a single off-diagonal element ($xy$) in the effective mass tensor leads to a coupling between
the $x$- and $y$-directions. 
The presence of the off-diagonal coupling additionally renormalizes the effective $xx$ and $yy$ spring constants
by a factor of $\mu_{xy}$. 
This may be expected from conservation of energy.
Because some of the quantum kinetic energy is transferred into the coupling between the $x$- and $y$-directions,
the $xx$ and $yy$ components of the kinetic energy must be correspondingly reduced to account
for this energy transfer into the additional degree of freedom.
Hence, the contribution from the off-diagonal coupling is opposite in sign to the diagonal coupling,
and the prefactor is twice the magnitude of that for a single diagonal term (half from $xx$ and half from $yy$). 
%

%
We now demonstrate the utility of our approach in the context of
simulating an exciton in an alkali halide crystal.
We model an electron-hole pair at a constant temperature of $T=300$~K,
and the electron and hole ring polymer each have $P=1024$ beads.
These MD simulations were performed using the LAMMPS software package~\cite{LAMMPS}
with Nos\'{e}-Hoover chains to maintain a constant temperature~\cite{Martyna1992}
and a Parrinello-Rahman barostat to maintain zero
pressure~\cite{shinoda2004rapid,parrinello1981polymorphic,Martyna:JCP:1994},
allowing the crystal to relax to the presence of the exciton. 
The ring polymers were massively thermostatted to ensure proper sampling
of the canonical distribution~\cite{Tuckerman:JCP:1993,Tuckerman_1996}.
We use the Tosi-Fumi model for KCl~\cite{Aragones_2012}.
Electrostatic interactions between charges of the same sign are described with a standard Coulomb potential,
while those between charges of opposite sign are described with a Shaw pseudopotential~\cite{Shaw},
with short-ranged cutoffs of 1.96~\AA, 1.75~\AA, and 1.69~\AA \ for electron-K$^+$,
hole-Cl$^-$, and electron-hole interactions. 
The electron-K$^+$ cutoff is that used by Parrinello and Rahman in their seminal study of F-centers
in KCl~\cite{Parrinello_1984}, the hole-Cl$^-$ cutoff corresponds to the Cl van der Waals radius,
and the electron-hole cutoff is crudely chosen to yield the band gap in the single bead limit;
the band gap is the energy difference between infinite separation and a perfectly overlapping electron and hole.
Further refinement of the latter cutoff can be performed, to match the exciton binding energy for example,
but we reserve this for future work and note that the cutoff used here yields reasonable predictions.
Long-ranged electrostatic interactions were evaluated using the particle-particle-particle mesh
Ewald method~\cite{pppm}.
Band structure calculations were performed using the GPAW software package~\cite{gpaw},
in combination with the atomic simulation environment (ASE)~\cite{ASE}, and employed the
PBE density functional approximation~\cite{pbe} with a plane-wave cutoff of 1200~eV and a
$12\times12\times12$ $k$-point mesh. 
Although the band gap is not properly described at this level of theory\cite{Perdew:IJQC:1985,Sanchez:PRL:2008}, 
the curvature of the bands near the gap is likely adequate, and we expect the effective mass for these materials
to be somewhat insensitive to the choice of semi-local functional. 
Note that high throughput calculations of more complex materials have also used DFT effective masses successfully 
in order to correlate transport and other properties from predicted band structures.\cite{Wang:PRX:2011,Hautier:NC:2013}
For highly accurate band structures and effective masses, 
approaches beyond semi-local DFT are required.\cite{Hybertsen:PRL:1985,Kim:PRB:2010,Olsen:NPJCM:2019}
Effective masses herein were determined using the effective mass calculator (EMC) program
using a five-point stencil to evaluate the second derivatives~\cite{emc}.
We find that $\mb^{-1}$ is approximately diagonal, such that
the masses of the electron and hole are approximately isotropic,
$m_{\rm \alpha \gamma}\approx  \delta_{\alpha\gamma}m^*$,
and equal to
$m_e^*=0.45 m_e$ and $m_h^*=5.2 m_e$, respectively, where $m_e$ is the bare
mass of an electron.
%

\begin{figure}[tb]
\begin{center}
\includegraphics[width=0.45\textwidth]{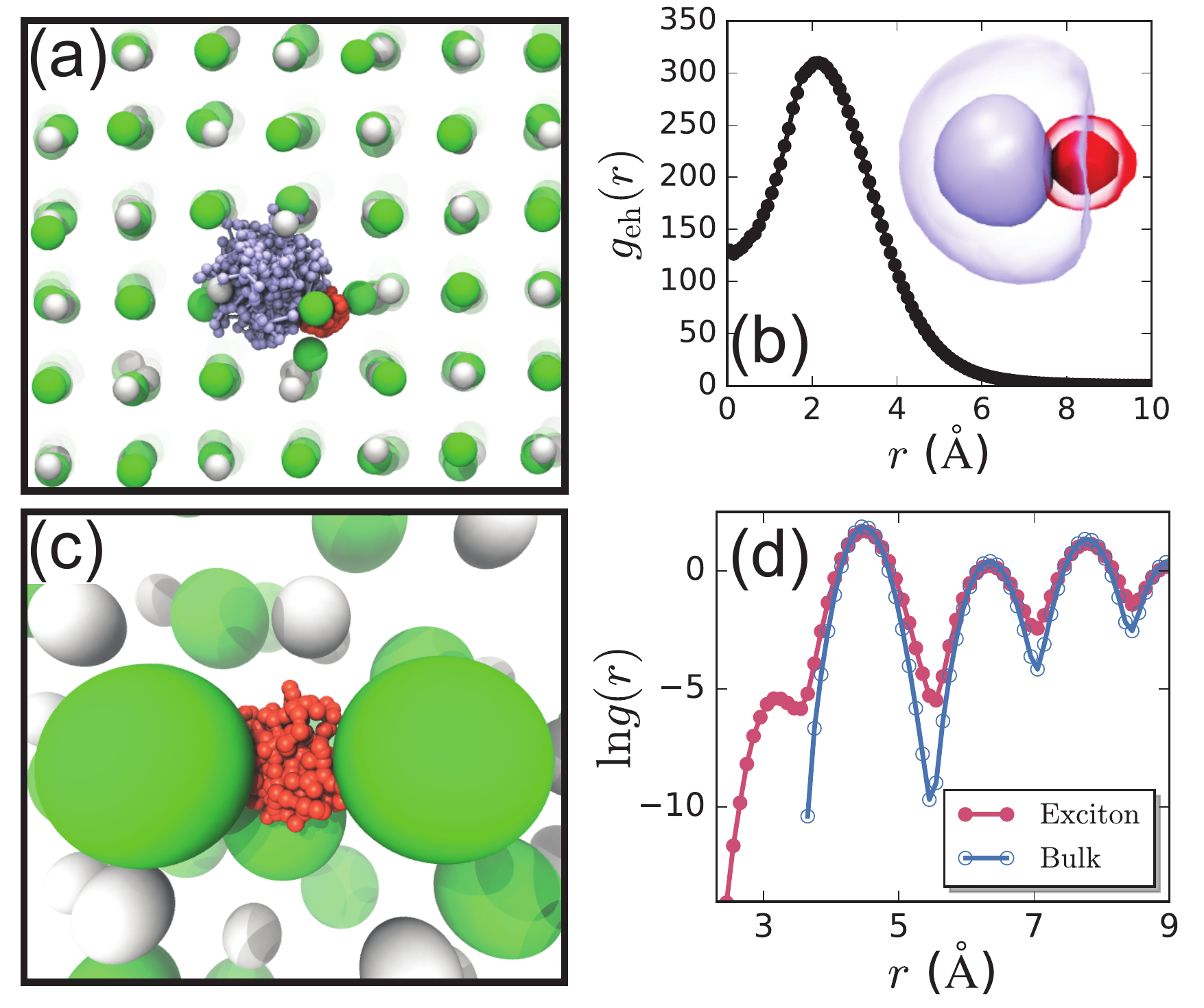}
\end{center}
\caption
{
(a) Snapshot of an electron (blue)-hole (red) pair --- an exciton --- in solid KCl from a single configuration of a
molecular dynamics simulation. K$^+$ and Cl$^-$ ions are drawn as white and green spheres. 
(b) The (bead-bead) pair distribution function, $g_{\rm eh}(r)$,
quantifies correlations between the electron and hole in solid KCl. 
The inset shows the electron (blue) and hole (red) three-dimensional spatial distribution function
computed in the rotating frame using the vector connecting the centroids as the $z$-axis,
indicating the formation of a dipolar exciton.
Solid and transparent isosurfaces are drawn to enclose approximately 90\% and 98\%
of the maximum density, respectively. 
(c) Simulation snaphot of the hole ring polymer localized between two chloride ions, forming a Cl$_2^-$ anion.
(d) Pair distribution function between chloride ions in a bulk KCl crystal (Bulk) and the KCl crystal with an
exciton present (Exciton). Note the appearance of a peak at low $r$ due to the formation of Cl$_2^-$
and similar states.
}
\label{fig:exah}
\end{figure}

%
A single exciton introduced into an otherwise
perfect alkali-halide crystal can self-trap and
create lattice defects~\cite{seitz1946color,seitz1954color,WSF_PRB86,Hess_2005,schwartz1997excitons,shluger1993small,WILLIAMS1990679}.
This self-trapping results in a structure resembling a closely separated F-center---H-center pair,
where the latter corresponds to a hole bridging two anions, Cl$_2^-$,
and similar states, \eg \ Cl$_3^{2-}$.
The self-trapped exciton is not spherical, as one might expect
using continuum theories of excitons in condensed phases.
Instead, the exciton is expected to be dipolar, with the
electron and hole separated by some average distance $R_{\rm eh}>0$
even in the bound, excitonic state~\cite{WSF_PRB86,Hess_2005,schwartz1997excitons}.
The formation of this self-trapped exciton state in solid KCl
is illustrated by the snapshot in Fig.~\ref{fig:exah}a.
The electron-hole pair distribution function
in Fig.~\ref{fig:exah}b demonstrates that
the dipolar exciton consists of an
electron and hole separated by $R_{\rm eh}\approx 2$~\AA.
This is further supported by the electron-hole spatial distribution function
shown in the inset of Fig.~\ref{fig:exah}b.
This is in very good agreement with previous interpretations of experiments
and detailed theoretical calculations
that also predict a dipolar exciton with $R_{\rm eh}\approx 2$~\AA~\cite{WSF_PRB86}.
While the light electron is delocalized over many ions (but localized with respect to a free electron
with the same effective mass),
the heavy hole is highly localized and bridges Cl$^-$ ions, as illustrated
by the snapshot in Fig.~\ref{fig:exah}c.
The Cl-Cl pair distribution functions, $g(r)$,
of the KCl crystal in the absence and presence of the exciton are compared
in Fig.~\ref{fig:exah}d, averaged over all Cl$^-$ in the system.
In the presence of the exciton, a peak at close Cl-Cl distances appears in $g(r)$, consistent with the
formation of Cl$_2^-$-like structures predicted in more detailed quantum calculations
and experiments~\cite{WSF_PRB86,Hess_2005,schwartz1997excitons,shluger1993small}.
We additionally note that the hole-anion interaction potential we employ is spherically
symmetric.
Including directionality into the hole-anion interactions, \eg \ through
the use of multisite ion models~\cite{Saxena:2015aa} for example,
may lead to even better descriptions of H-centers.
%


\begin{figure}[tb]
\begin{center}
\includegraphics[width=0.34\textwidth]{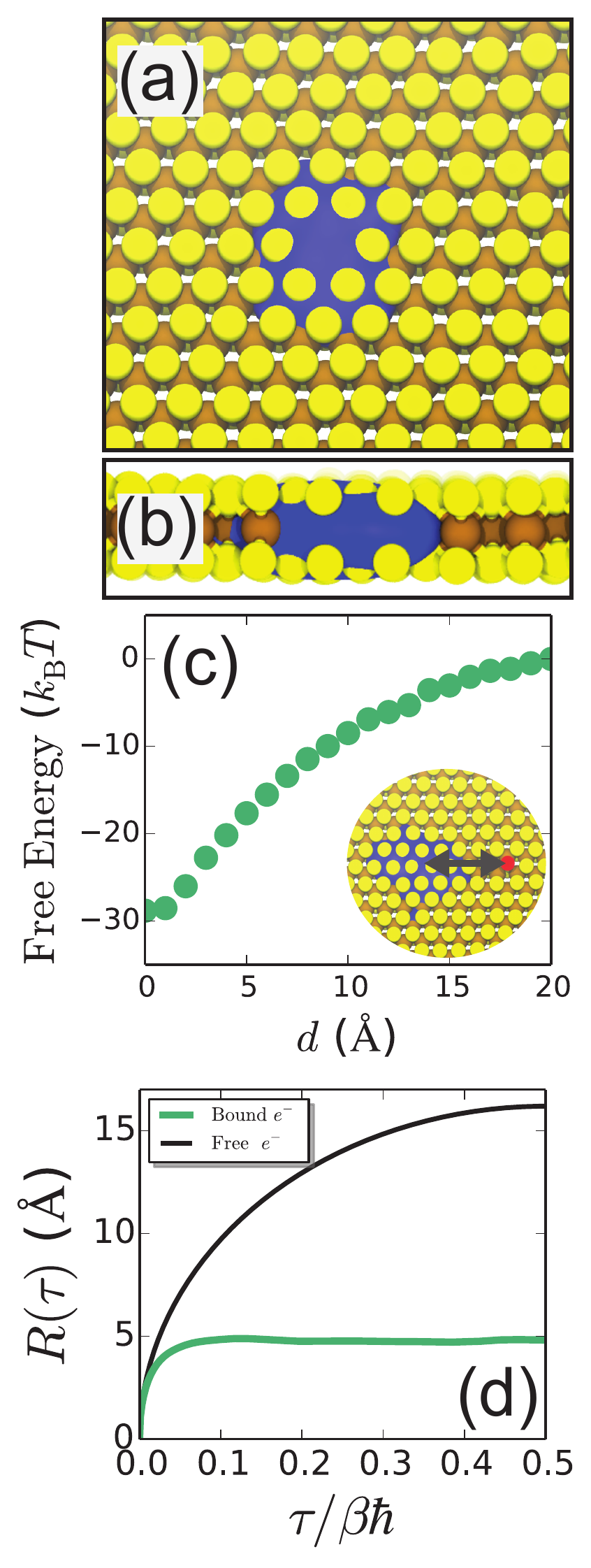}
\end{center}
\caption
{
Snapshots showing (a) top and (b) side views of an electron (blue surface) in monolayer MoS$_2$ (tan Mo and yellow S).
(c) Free energy as a function of the electron--S-vacancy in-plane distance.
Inset shows an illustration of the electron--S-vacancy distance, $d$; the vacancy is shown as a red circle. 
(d) Imaginary time root mean squared displacement, $\mathcal{R}(\tau)$, for a free effective mass electron
in two dimensions and
an electron bound to the S-vacancy.
}
\label{fig:mos2}
\end{figure}

%
The EMPI formalism can also describe
the effects of reduced dimensionality.
For example, monolayer transition metal dichalcogenides, such as MoS$_2$,
are two-dimensional materials that exhibit quantum confinement in the two-dimensional plane of the lattice~\cite{Wang_2012,Chhowalla_2013,Berkelbach_2013,Wang_2018,Berkelbach_Review}.
In this case, the effective masses suggest that the quasiparticles are essentially confined to the $xy$-plane,
$m_{xx}=m_{yy}\approx0.562 m_e <<m_{zz}\approx960$.
The large $z$-component of the effective mass tensor manifests as a high spring constant
$\kappa_{zz}$ that confines the quasiparticle significantly in the $z$-direction,
while allowing the particle to spread in the two-dimensional ($xy$) plane,
as illustrated by the snapshots in Fig.~\ref{fig:mos2}a,b.
We demonstrate the utility of EMPIMD simulations for systems with reduced dimensionality by studying
the binding of an excess electron in monolayer MoS$_2$ to a sulfur vacancy, which carries an effective positive charge.
Monolayer MoS$_2$ is modeled using the Stillinger-Weber potential designed to capture the
structure and vibrational properties of the monolayer~\cite{MoS2Model,remsing2017thermal}, which readily
enables modeling of defects. 
The interactions between the electron and the S atoms are modeled using a Coulomb potential,
and those between the electron and the Mo atoms are modeled using a Shaw pseudopotential~\cite{Shaw}
with a short-ranged cutoff of 0.18~\AA. 
The partial charges on the Mo and S atoms, used for interactions with the electron,
are those of Sresht \etal~\cite{Sresht:2017aa}.
Simulations are performed using LAMMPS~\cite{LAMMPS} with appropriate
Nos\'{e}-Hoover thermostatting to maintain a constant temperature of 300~K.
The anisotropic spring constants of the electron ring polymer are incorporated
using PLUMED~\cite{PLUMED}.
The electron---S-vacancy potential of mean force is calculated using umbrella sampling~\cite{UmbrellaSampling}
combined with UWHAM~\cite{UWHAM,MBAR}, where we biased the two-dimensional ($xy)$ distance between the electron
and the location of the vacancy using harmonic potentials in PLUMED~\cite{PLUMED}.

Spatially-resolved photoluminescence (PL) spectroscopy has discovered that excitonic hot spots appear
at the location of sulfur vacancies in monolayer MoS$_2$~\cite{Chow:ACSNano:2015,Tongay_2013,Li_2014,Amani1065}.
The higher intensity peaks in the vicinity of sulfur vacancies suggest that excess electrons in doped MoS$_2$
are bound to these vacancies.
Increasing the concentration of sulfur vacancies
results in the appearance of a new, lower energy peak in the PL spectra,
further suggesting the validity of this interpretation.
Our EMPIMD simulations further support this picture of strong
electron---S-vacancy interactions in monolayer MoS$_2$ leading to the formation of a trap state.
To quantify the interactions between the electron and a sulfur vacancy,
the free energy as a function of electron-vacancy distance
is shown in Fig.~\ref{fig:mos2}c.
We find a binding free energy of approximately 30$\kT$ at 300~K, in agreement
with the range of energies predicted by kinetic modeling of spectroscopic measurements~\cite{Goodman_2017}.
This binding free energy is also in good agreement
with the energy difference between the trap state and the conduction band
predicted by density functional theory (DFT) calculations~\cite{Qiu_2013}.
Our approach additionally enables the characterization of the effects of the sulfur vacancy on the electron.
For example, examination of the imaginary time root mean squared displacement~\cite{Chandler:AnnuRevPhysChem:1994,Miller_2008}, $\mathcal{R}(\tau)=\avg{ \len{\rb_{xy}(\tau)-\rb_{xy}(0)}^2}^{1/2}$,
where $\rb_{xy}$ indicates that distances in-plane were considered in the calculation.
Comparison of $\mathcal{R}(\tau)$ for the trapped electron with that expected for a free electron in MoS$_2$,
$\mathcal{R}_{\rm free}(\tau)=\brac{2\lambda^2 (t/\beta\hbar)(1-t/\beta\hbar)}^{1/2}$,
where $\lambda^2=\beta\hbar^2/m$ and $m=m_{xx}=m_{yy}$,
indicates that binding to the sulfur vacancy traps the excess electron, as shown in Fig.~\ref{fig:mos2}d.
The value of $\mathcal{R}(\beta\hbar/2)\approx5$~\AA \ yields an estimate for the effective size of the electron in good agreement
with that determined for the trap state from DFT calculations~\cite{Qiu_2013}.
Moreover, writing $\mathcal{R}(\tau)$ in the basis of electronic eigenstates~\cite{Chandler:AnnuRevPhysChem:1994,Miller_2008},
\begin{equation}
\mathcal{R}^2(\tau) = \frac{4}{Z} \sum_{n,m} e^{-\beta E_n} x_{nm}^2 (1- e^{-\tau(E_m-E_n)/\hbar}),
\label{eq:ee}
\end{equation}
where we have considered the 2D case assuming $x$ and $y$ are equivalent by symmetry, $Z$ is the partition function,
$x_{nm}=\mel{n}{x}{m}$, and $E_n$ is the energy of eigenstate $n$,
further suggests that the electron is in a deep trap state.
In order for $\mathcal{R}(\tau)$ to be independent of $\tau$,
as is the case for $\tau\gg0$, Eq.~\ref{eq:ee} must be dominated
by a single, localized eigenstate with a significant gap to the first excited state. 
This ground state that dominates the behavior of the vacancy-bound electron is the trap state.
In contrast, the spatially extended states sampled by the free electron
correspond to thermally accessible excited electronic states.
These results highlight the utility of EMPIMD simulations in quantifying the thermodynamics 
of quasiparticle interactions in anisotropic materials. 
%

%
In this Communication, we have presented a formulation of effective mass path integral simulations.
This approach incorporates effective mass theory into the path integral description of
electrons and holes for their simulation in condensed phases,
extending previous PI-based methods using scalar effective masses
to materials with anisotropic electronic structure.
We expect that this approach will find wide use in a variety of applications in chemical and materials physics,
including the simulation of exciton and charge carrier structure and dynamics in materials.
We note that our implementation of the EMPIMD simulations described above has not exploited
the many significant advances made in the context of path-integral simulations in recent years.
However, the EMPIMD method is readily amenable to such approaches,
including novel integration schemes~\cite{Venkat:JCP:2016,Miller:JCP:2019,Poltavsky:2020aa,Roman:JCP:2020},
thermostats~\cite{Ceriotti_2011,Rossi:JCP:2014,Rossi:JCP:2018},
and nearly all other developments in the context of path integral and ring-polymer techniques~\cite{Markland:2008,MARKLAND2008256,Menzeleev_2010,Richardson_2013,Ananth_2013,Menzeleev_2014,Duke_2015}. 
Of additional importance for the future of EMPI models is the development of accurate electron-environment and
hole-environment pseudopotentials, beyond the simplistic, spherically-symmetric charge-charge pseudopotentials
used here~\cite{mayer2004band,zhang2018light}, as well as the inclusion
of modified electrostatics due to dielectric screening
in low-dimensional materials~\cite{Berkelbach_2013,Cho_2018}.
These effective interaction potentials will impact the accuracy of the predictions made by EMPI approaches.
Finally, we note that this framework can also be readily used within the
ring-polymer MD approximation and similar
approaches~\cite{Craig_2004,Hone_2006,Habershon:2013aa,Welsch_2016}
to model the quantum dynamics of quasiparticles in complex environments,
and future work will focus on extending these approaches to EMPI simulations.
\\

\acknowledgments
RCR acknowledges the Office of Advanced Research Computing (OARC) at Rutgers,
The State University of New Jersey
for providing access to the Amarel cluster
and associated research computing resources that have contributed to the results reported here.
JEB was supported by start-up funds provided by
Appalachian State University.
We thank Axel Kohlmeyer for helpful discussions regarding the LAMMPS code. 


\bibliography{Refs}

\end{document}